\documentclass[sigconf]{acmart}

\usepackage{xspace}
\usepackage{paralist}
\usepackage{blindtext}
\usepackage{url}
\usepackage{hyperref}
\usepackage{listings}
\usepackage{xcolor}
\usepackage{booktabs}
\usepackage{pgfplots}
\usepackage{tikz}

\pgfplotsset{width=7cm,compat=1.18}

\newcommand{\beamng}{\textsc{BeamNG.tech}\xspace}
\newcommand{\beamngdrive}{\textsc{BeamNG.drive}\xspace}
\newcommand{\beamnggmbh}{\textsc{BeamNG GmbH}\xspace}
\newcommand{\mongodb}{\textsc{MongoDB}\xspace}
\newcommand{\frenetic}{\textsc{Frenetic}\xspace}
\newcommand{\freneticv}{\textsc{FreneticV}\xspace}
\newcommand{\ambiegen}{\textsc{AmbieGen}\xspace}
\newcommand{\docker}{\textsc{Docker}\xspace}
\newcommand{\dataset}{\textsc{SensoDat}\xspace}

\colorlet{punct}{red!60!black}
\definecolor{background}{HTML}{EEEEEE}
\definecolor{delim}{RGB}{20,105,176}
\colorlet{numb}{magenta!60!black}

\lstdefinelanguage{json}{
    basicstyle=\normalfont\ttfamily,
    numbers=left,
    numberstyle=\scriptsize,
    stepnumber=1,
    numbersep=8pt,
    showstringspaces=false,
    breaklines=true,
    frame=lines,
    backgroundcolor=\color{background},
    literate=
     *{0}{{{\color{numb}0}}}{1}
      {1}{{{\color{numb}1}}}{1}
      {2}{{{\color{numb}2}}}{1}
      {3}{{{\color{numb}3}}}{1}
      {4}{{{\color{numb}4}}}{1}
      {5}{{{\color{numb}5}}}{1}
      {6}{{{\color{numb}6}}}{1}
      {7}{{{\color{numb}7}}}{1}
      {8}{{{\color{numb}8}}}{1}
      {9}{{{\color{numb}9}}}{1}
      {:}{{{\color{punct}{:}}}}{1}
      {,}{{{\color{punct}{,}}}}{1}
      {\{}{{{\color{delim}{\{}}}}{1}
      {\}}{{{\color{delim}{\}}}}}{1}
      {[}{{{\color{delim}{[}}}}{1}
      {]}{{{\color{delim}{]}}}}{1},
}

\acmConference[MSR 2024]{21st International Conference on Mining Software Repositories}{April 2024}{Lisbon, Portugal}

\usepackage{graphicx} 

\title{SensoDat: Simulation-based Sensor Dataset of Self-driving Cars}

\author{Christian Birchler}
\email{christian.birchler@{zhaw,unibe}.ch}
\affiliation{
  \institution{Zurich University of Applied Sciences \\ University of Bern}
  \country{Switzerland}
}

\author{Cyrill Rohrbach}
\email{cyrill.rohrbach@students.unibe.ch}
\affiliation{
  \institution{University of Bern}
  \country{Switzerland}
}

\author{Timo Kehrer}
\email{timo.kehrer@unibe.ch}
\affiliation{
  \institution{University of Bern}
  \country{Switzerland}
}

\author{Sebastiano Panichella}
\email{sebastiano.panichella@zhaw.ch}
\affiliation{
  \institution{Zurich University of Applied Sciences}
  \country{Switzerland}
}

\begin{abstract}
Developing tools in the context of autonomous systems \cite{SBFT-UAV2024,khatiri2023simulation} such as self-driving cars (SDCs) is time-consuming and costly since researchers and practitioners rely on expensive computing hardware and simulation software.
We propose \dataset, a dataset of 32,580 executed simulation-based SDC test cases generated with state-of-the-art test generators for SDCs.
The dataset consist of trajectory logs and a variety of sensor data from the SDCs (e.g., rpm, wheel speed, brake thermals, transmission, etc.) represented as a time series.
In total, \dataset provides data from 81 different simulated sensors.
Future research in the domain of SDCs does not necessarily depend on executing expensive test cases when using \dataset.
Furthermore, with the high amount and variety of sensor data, we think \dataset can contribute to research, particularly for AI development, regression testing techniques for simulation-based SDC testing, flakiness in simulation, etc.
\\
Link to the dataset: \url{https://doi.org/10.5281/zenodo.10307479}

\end{abstract}

\begin{document}

\maketitle

\section{Introduction}
Testing self-driving cars (SDCs) is crucial for maintaining high security levels and minimizing potential threats to humans.
While infield SDC testing is costly, simulation technologies offer a safer alternative.
However, conducting simulation-based tests demands increased computational resources, particularly GPUs for accelerated computation of physical dynamics~\cite{DBLP:journals/tse/ManciniMT23,DBLP:conf/gecco/ArrietaWAMSE18}.


To address the challenge of minimizing the costs of simulation-based testing, recent research focused on various regression testing techniques for simulation-based tests.
Those techniques aim to test SDCs cost-effectively while maintaining the system's safety.
For instance, by test prioritization~\cite{DBLP:journals/jss/ArrietaWSE19,DBLP:journals/tosem/BirchlerKDPP23}, in which the tests of the test suite are prioritized, i.e., sorted in a way so that the testing phase reveals faults of the system earlier.

However, to conduct research on optimizing simulation-based testing, researchers rely on expensive test executions in simulation environments~\cite{DBLP:journals/tse/ManciniMT23,DBLP:conf/gecco/ArrietaWAMSE18}.
These additional computational costs are mainly due to the expensive computation simulating the physics of the environment, which is not the case when testing traditional software systems.
Thus, those tests are expensive and often not affordable when a large amount of data is required, e.g., for ML and DNNs.

To overcome the issue of running simulations to obtain the execution data of simulations, researchers can use existing datasets of executed simulation-based tests.
There are only a few datasets available consisting of simulation data for SDCs~\cite{DBLP:confMSRr/LuYA23,prioritizer-zenodo}.
Despite the existence of those datasets, in most cases, the simulators used are not maintained anymore.
Popular simulators like Udacity~\cite{udacity}, Apollo~\cite{apollo}, SVL~\cite{svl}, and DeepDrive~\cite{deepdrive}, were used in the past for research purposes but unfortunately, the active development of these simulators has been stopped by the maintainers or have long release cycles.
%

We propose a dataset consisting of simulation data of executed SDC test cases in the \beamng simulation environment.
The \beamng simulator is known in academia~\cite{DBLP:conf/kbse/BirchlerRKGLHKP23,DBLP:journals/ese/BirchlerKBGP23,DBLP:conf/sbst/PanichellaGZR21,DBLP:conf/sbst/GambiJRZ22,DBLP:conf/icse/BiagiolaKPR23,10.1145/3544792} and is based on the popular \beamngdrive game, which is actively developed and maintained by \beamnggmbh.
The availability of simulation data of SDCs enhances the research on SDC testing in simulation.
Researchers and practitioners do not rely on executing expensive test cases to develop and evaluate regression testing techniques.
Having the dataset publicly available eases the research for the domain of SDC software, especially for researchers and practitioners who can not afford expensive computing hardware.
Furthermore, the availability of an open dataset improves the reproducibility and comparability of research results in various areas.

\begin{figure}
    \centering
    \includegraphics[width=\linewidth]{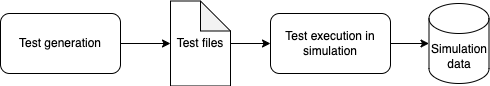}
    \caption{High-level dataset generation process}
    \label{fig:methodology-overview}
\end{figure}

\section{Methodology}
As illustrated in Figure~\ref{fig:methodology-overview}, to start a data collection process of sensor data of simulation-based SDC test cases, we have to generate test cases with different test generators (Section~\ref{sec:test-generators}).
After the generation of test cases, we execute those test cases in a simulation environment (Section~\ref{sec:simulation-platform}).
During the test execution, we collect the data from various simulated sensors of the SDC.

\subsection{Test generation}\label{sec:test-generators}
We generate test cases that are based on three different test generators for the \beamng simulator.
All of these test generators were developed in the context of tool competitions at the SBST~\cite{DBLP:conf/sbst/PanichellaGZR21,DBLP:conf/sbst/GambiJRZ22} and SBFT~\cite{DBLP:conf/icse/BiagiolaKPR23} workshop.

\begin{lstlisting}[language={json}, numbers={left}, basicstyle={\tiny\ttfamily}, backgroundcolor={}, captionpos={b}, caption={Metadata of a test execution consisting of simulation configuration and test outcome}, label={lst:test-metadata}]
{
  "_id": {...},
  "OpenDRIVE": {
    "header": {
      "@name": "mt_set5/frenetic/test-440",
      "sdc_test_info": {
        "@test_id": "mt_set5/frenetic/test-440",
        "@test_outcome": "PASS",
        "@predicted_test_outcome": "null",
        "@test_duration": "11.067427158355713",
        "@is_valid": "True"
      },
      "run_config": {
        "@rf": "1.5",
        "@oob": "0.5",
        "@max_speed": "120",
        "@obstacles": "False",
        "@bump_dist": "20",
        "@delineator_dist": "None",
        "@tree_dist": "None",
        "@field_of_view": "120"
      }
    },
    "road": {...}
  },
  "execution_data": {...}
}
\end{lstlisting}

\subsubsection{\frenetic}
The \frenetic tool~\cite{DBLP:conf/sbst/CastellanoCTKZA21} uses a genetic algorithm to minimize the distance between the SDC and the edge of the road.
For its computation, it leverages the concepts of frenet frames and curvature-based road representations.
Thus, it converts the solutions back to the Cartesian space, which is required for the simulator.

\subsubsection{\freneticv}
The \freneticv tool~\cite{DBLP:conf/sbst/CastellanoKCA22} is an extension of \frenetic, which is able to reduce the amount of invalid roads.
An invalid road is for example a road which has overly sharp turns, or if it intersects with itself.
This definition of road validity is given by the SBST tool competition~\cite{DBLP:conf/sbst/GambiJRZ22}, for which the tool was developed for.

\subsubsection{\ambiegen}
The \ambiegen~\cite{DBLP:conf/sbst/HumeniukAK22} tool is a test generator that uses a multi-objective approach using NSGA-II~\cite{DBLP:conf/ppsn/DebAPM00}.
Using the diversity preservation and the fault revealing power as the objectives, \ambiegen generates test cases that are more likely to reveal faults based on the OOB metric.

\begin{table}[t]
    \centering
    \scriptsize
    \caption{Database storage size}
    \begin{tabular}{lcc} \toprule
        \textbf{Collection} & \textbf{\# Documents} & \textbf{Storage size} \\ \midrule
        campaign\_2\_ambiegen & 973 & 108.36 MB \\
        campaign\_2\_frenetic & 928 & 109.46 MB \\
        campaign\_2\_frenetic\_v & 944 & 41.93 MB \\
        campaign\_3\_ambiegen & 964 & 109.85 MB \\
        campaign\_3\_frenetic & 954 & 112.73 MB \\
        campaign\_4\_ambiegen & 965 & 111.98 MB \\
        campaign\_4\_frenetic & 964 & 113.87 MB \\
        campaign\_4\_frenetic\_v & 525 & 63.43 MB \\
        campaign\_5\_ambiegen & 958 & 109.59 MB \\
        campaign\_5\_frenetic & 945 & 112.29 MB \\
        campaign\_5\_frenetic\_v & 940 & 112.81 MB \\
        campaign\_6\_ambiegen & 959 & 111.76 MB \\
        campaign\_6\_frenetic & 944 & 111.11 MB \\
        campaign\_6\_frenetic\_v & 764 & 91.14 MB \\
        campaign\_7\_ambiegen & 963 & 110.00 MB \\
        campaign\_7\_frenetic & 967 & 114.09 MB \\
        campaign\_7\_frenetic\_v & 47 & 5.67 MB \\
        campaign\_8\_ambiegen & 952 & 110.76 MB \\
        campaign\_8\_frenetic & 952 & 112.25 MB \\
        campaign\_9\_ambiegen & 953 & 109.20 MB \\
        campaign\_9\_frenetic & 964 & 113.57 MB \\
        campaign\_10\_ambiegen & 971 & 63.95 MB \\
        campaign\_11\_ambiegen & 973 & 72.79 MB \\
        campaign\_11\_frenetic & 866 & 66.44 MB \\
        campaign\_11\_frenetic\_v & 953 & 73.52 MB \\
        campaign\_12\_frenetic & 956 & 110.11 MB \\
        campaign\_12\_freneticV & 942 & 114.00 MB \\
        campaign\_13\_ambiegen & 954 & 68.83 MB \\
        campaign\_13\_frenetic & 959 & 72.52 MB \\
        campaign\_13\_frenetic\_v & 951 & 71.48 MB \\
        campaign\_14\_ambiegen & 959 & 70.14 MB \\
        campaign\_14\_frenetic & 866 & 64.05 MB \\
        campaign\_14\_frenetic\_v & 934 & 70.16 MB \\
        campaign\_15\_ambiegen & 952 & 110.50 MB \\
        campaign\_15\_frenetic & 870 & 102.61 MB \\
        campaign\_15\_freneticV & 949 & 114.67 MB \\ \midrule
        \textbf{Total} & \textbf{32,580} & \textbf{3.34 GB} \\ \bottomrule
    \end{tabular}
    \label{tab:overview-db-collections-and-storage-sizzes}
\end{table}

\begin{table*}[t]
    \centering
    \scriptsize
    \caption{Overview of available types of sensor data}
    \resizebox{\linewidth}{!}{
    \begin{tabular}{l|l|l|l|l|l|l|l|l}\toprule
    fuel                &   steering\_input           &  oil                        & exhaust\_flow         & 4x brakeCoreTemperature             & left\_signal            & airspeed              & brake\_input              & signal\_l         \\
    low fuel            &   rpm spin                  &  lowhighbeam                & fog\_lights           & 4x brakeThermalEfficiency           & signal\_r               & abs                   & tcs                       & parking           \\   
    gear                &   airflow speed             &  lowbeam                    & fuel\_volume          & 4x brakeSurfaceTemperature          & right\_signal           & steering              & ignition                  & hazard            \\ 
    odometer            &   lights                    &  high beam                  & fuel\_capacity        & engineRunning                       & tcs\_active             & isYCBrakeActive       & gearboxMode               & clutch\_input     \\
    brake               &   horn                      &  brakelight\_signal\_R      & gear\_a               & running                             & water\_temperature      & isTCBrakeActive       & lightbar                  & abs\_active       \\
    throttle            &   hasABS                    &  brakelight\_signal\_L      & gear\_index           & low-pressure                        & brake\_lights           & driveshaft            & headlights                & engine\_throttle  \\            
    parking brake       &   altitude                  &  lowhighbeam\_signal\_R     & gear\_m               & rpm                                 & check\_engine           & wheelspeed            & oil\_temperature          & esc\_active       \\  
    throttle\_input     &   dseColor                  &  lowhighbeam\_signal\_L     & hazard\_signal        & clutch                              & clutch\_ratio           & esc                   & radiator\_fan\_spin       & avg\_wheel\_av    \\
    reverse             &   virtualAirspeed           &  turn signal                & is\_shifting          & parkingbrake\_input                 & engine\_load            & smoothShiftLogicAV    & rpm\_tacho                & freezeState       \\ \bottomrule
    \end{tabular}}
    \label{tab:overview-of-sensors}
\end{table*}

\subsection{Simulation platform}\label{sec:simulation-platform}
We used the \beamng~\cite{beamng_tech} simulator with \textsc{SDC-Scissor}~\cite{DBLP:conf/wcre/BirchlerGKGP22} to execute the generated test cases.
\beamng is a widely used simulator for researching testing of SDCs in academia~\cite{DBLP:conf/sbst/PanichellaGZR21,DBLP:conf/sbst/GambiJRZ22,DBLP:conf/icse/BiagiolaKPR23}.
It is a high-fidelity soft-body physics simulator that accurately simulates the car's dynamics, such as the pressure and deformation on the chassis and engine, clutch, and its related components.
We conjecture obtaining as realistic data as possible using the \beamng simulator.

The vendor of \beamng provides researchers with a free academic license, which makes it highly accessible for researchers.
Furthermore, the simulator provides a Python API to control the simulation process, including defining the environment with the road shape and gathering sensor data.

\subsection{Data generation}
\subsubsection{Process}
We generated data by conducting 14 simulation campaigns.
Every campaign follows the process as illustrated in Figure~\ref{fig:methodology-overview}.
We generate test cases with test generators (Section~\ref{sec:test-generators}).
After the generation, we execute the test cases in the simulation environment and collect the sensor data at runtime.
Furthermore, we collect the information if the SDC violated the OOB safety metric and, therefore, passes or fails the test case.

\subsubsection{Infrastructure}
The simulations were executed on a machine with the following hardware specifications:
\textit{AMD Ryzen 7 3800X} 8 core 16 threads, 64 GB DDR4 RAM, \textit{NVIDIA GeForce GTX 1080} 8GB, \textit{Windows 11}.

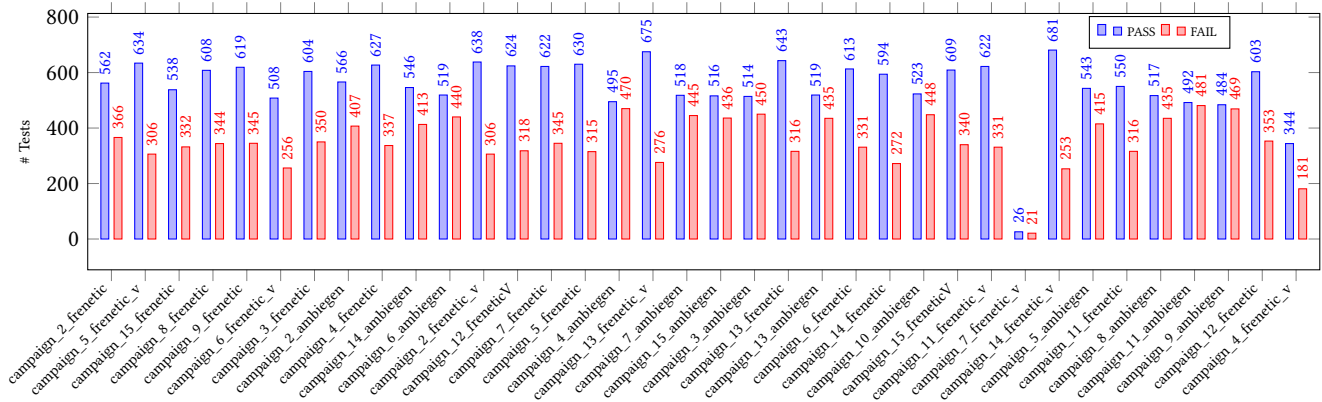
\begin{figure*}[t]
    \centering
    \resizebox{\linewidth}{!}{
\begin{tikzpicture}
\begin{axis}[
ybar,
width=18cm,
height=5cm,
bar width=3pt,
enlargelimits=0.02,
enlarge y limits=0.2,
legend style={at={(0.87, 0.99), font=\tiny},
anchor=north,legend columns=-1},
ylabel={\scriptsize \# Tests},
symbolic x coords={campaign\_2\_frenetic, campaign\_5\_frenetic\_v, campaign\_15\_frenetic, campaign\_8\_frenetic, campaign\_9\_frenetic, campaign\_6\_frenetic\_v, campaign\_3\_frenetic, campaign\_2\_ambiegen, campaign\_4\_frenetic, campaign\_14\_ambiegen, campaign\_6\_ambiegen, campaign\_2\_frenetic\_v, campaign\_12\_freneticV, campaign\_7\_frenetic, campaign\_5\_frenetic, campaign\_4\_ambiegen, campaign\_13\_frenetic\_v, campaign\_7\_ambiegen, campaign\_15\_ambiegen, campaign\_3\_ambiegen, campaign\_13\_frenetic, campaign\_13\_ambiegen, campaign\_6\_frenetic, campaign\_14\_frenetic, campaign\_10\_ambiegen, campaign\_15\_freneticV, campaign\_11\_frenetic\_v, campaign\_7\_frenetic\_v, campaign\_14\_frenetic\_v, campaign\_5\_ambiegen, campaign\_11\_frenetic, campaign\_8\_ambiegen, campaign\_11\_ambiegen, campaign\_9\_ambiegen, campaign\_12\_frenetic, campaign\_4\_frenetic\_v},
xtick=data,
x tick label style={rotate=45, anchor=east, font=\scriptsize},
nodes near coords,
nodes near coords align={vertical},
nodes near coords style={
        anchor=west,
        rotate=90,
        font=\scriptsize,
        #1
    }
]

\addplot coordinates {(campaign\_2\_frenetic,562) (campaign\_3\_frenetic,604) (campaign\_4\_frenetic,627) (campaign\_5\_frenetic,630) (campaign\_6\_frenetic,613) (campaign\_7\_frenetic,622) (campaign\_8\_frenetic,608) (campaign\_9\_frenetic,619) (campaign\_11\_frenetic,550) (campaign\_12\_frenetic,603) (campaign\_13\_frenetic,643) (campaign\_14\_frenetic,594) (campaign\_15\_frenetic,538)
(campaign\_2\_frenetic\_v,638) (campaign\_4\_frenetic\_v,344) (campaign\_5\_frenetic\_v,634) (campaign\_6\_frenetic\_v,508) (campaign\_7\_frenetic\_v,26) (campaign\_11\_frenetic\_v,622) (campaign\_12\_freneticV,624) (campaign\_13\_frenetic\_v,675) (campaign\_14\_frenetic\_v,681) (campaign\_15\_freneticV,609)
(campaign\_2\_ambiegen,566) (campaign\_3\_ambiegen,514) (campaign\_4\_ambiegen,495) (campaign\_5\_ambiegen,543) (campaign\_6\_ambiegen,519) (campaign\_7\_ambiegen,518) (campaign\_8\_ambiegen,517) (campaign\_9\_ambiegen,484) (campaign\_10\_ambiegen,523) (campaign\_11\_ambiegen,492) (campaign\_13\_ambiegen,519) (campaign\_14\_ambiegen,546) (campaign\_15\_ambiegen,516)};

\addplot coordinates {(campaign\_2\_frenetic,366) (campaign\_3\_frenetic,350) (campaign\_4\_frenetic,337) (campaign\_5\_frenetic,315) (campaign\_6\_frenetic,331) (campaign\_7\_frenetic,345) (campaign\_8\_frenetic,344) (campaign\_9\_frenetic,345) (campaign\_11\_frenetic,316) (campaign\_12\_frenetic,353) (campaign\_13\_frenetic,316) (campaign\_14\_frenetic,272) (campaign\_15\_frenetic,332)
(campaign\_2\_frenetic\_v,306) (campaign\_4\_frenetic\_v,181) (campaign\_5\_frenetic\_v,306) (campaign\_6\_frenetic\_v,256) (campaign\_7\_frenetic\_v,21) (campaign\_11\_frenetic\_v,331) (campaign\_12\_freneticV,318) (campaign\_13\_frenetic\_v,276) (campaign\_14\_frenetic\_v,253) (campaign\_15\_freneticV,340)
(campaign\_2\_ambiegen,407) (campaign\_3\_ambiegen,450) (campaign\_4\_ambiegen,470) (campaign\_5\_ambiegen,415) (campaign\_6\_ambiegen,440) (campaign\_7\_ambiegen,445) (campaign\_8\_ambiegen,435) (campaign\_9\_ambiegen,469) (campaign\_10\_ambiegen,448) (campaign\_11\_ambiegen,481) (campaign\_13\_ambiegen,435) (campaign\_14\_ambiegen,413) (campaign\_15\_ambiegen,436)};

\legend{PASS,FAIL}
\end{axis}
\end{tikzpicture}}
    \caption{Test outcome distribution among \mongodb collections}
    \label{fig:test-outcome-dictribution}
\end{figure*}

\section{Data storage}
We use \mongodb as database technology to efficiently query and perform analyses on the data.
Since the raw data obtained from the simulations is mainly stored as JSON files, the mapping of JSON files to documents in \mongodb is straightforward
For each simulation campaign, a dedicated collection is created in the database.

\subsection{Database setup}
First, to set up the database, download the raw data and code from the public repository as described in Section~\ref{sec:data-availability}.
Furthermore, it is required to have \textsc{Docker}~\cite{docker} installed on the machine.

\begin{lstlisting}[language={}, numbers={left}, basicstyle={\scriptsize\ttfamily}, backgroundcolor={}, captionpos={b}, caption={Database setup with \docker}]
docker-compose -f ./environment/docker-compose.yml up -d --build
docker ps # Verify if container is up and running
docker cp ./data uploader:/app/data
docker exec uploader unzip ./data/data.zip -d ./data
docker cp ./code uploader:/app/code
docker exec -it uploader python ./code/fill_mongodb.py
\end{lstlisting}

A locally deployed \mongodb instance will provide all simulation data by following the aforementioned instructions.
The data is accessible by using a compatible \mongodb client (e.g., \textsc{MongoDB Compass}~\cite{mongodb-compass}) and the default credentials (\textit{username}: \texttt{msr}, \textit{password}: \texttt{fooBar}).

\subsection{Database query}
Given the database is up and running, querying the data is simple.
The queries must conform to the \mongodb language specification as illustrated in a sample query in Listing~\ref{lst:sample-query}.
In this example, we use \textsc{pymongo}~\cite{pymongo} as a client library to connect to \mongodb for the \textsc{Python} programming language.
The query operates on the \texttt{campaign\_2\_frenetic} database collection and counts the number of passing test executions.
For this specific query, we make use of the document structure as illustrated in Listing~\ref{lst:test-metadata}.

\newpage
\begin{lstlisting}[language={python}, numbers={left}, basicstyle={\scriptsize\ttfamily}, backgroundcolor={}, captionpos={b}, caption={Sample query with \textsc{pymongo}}, label={lst:sample-query}]
from pymongo.mongo_client import MongoClient
from pymongo.server_api import ServerApi

uri = "mongodb://msr:fooBar@localhost:27017/?authMechanism=DEFAULT"
client = MongoClient(uri, server_api=ServerApi('1'))
db = client.get_database('sdc_sim_data')

query_passing = \
    {"OpenDRIVE.header.sdc_test_info.@test_outcome": "PASS"}
nr_passing_tests = db['campaign_2_frenetic'] \
    .count_documents(filter=query_passing)
print(nr_passing_tests)
\end{lstlisting}

Adjusting the query string, a user can also query for specific sensor values.
For example, the user can query data for a time range before a fault has occurred in the simulation.
For this, the user needs to identify the location of the desired field in the database document and write the appropriate query string.

\section{Data charactaristics}
Table~\ref{tab:overview-db-collections-and-storage-sizzes} overviews the total amount of collections and their corresponding required storage sizes.
From the simulator in which we conducted 14 test campaigns, we obtained total data from 32,580 test executions.
Figure~\ref{fig:test-outcome-dictribution} depicts the distribution of failing and passing test executions among 14 test campaigns.
In total, we have 19,926 passing and 12,654 failing test executions.
The distribution depends on the out-of-bound (OOB) metric, which acts as the oracle.
For this dataset, we set $OOB=0.5$, in order to have a fairly balanced dataset of passing and failing tests.

Next to the metadata (see Listing~\ref{lst:test-metadata}), we obtain data from the simulation as time series.
For each data field, we obtain its state annotated with the timestamp.
The timestamp reflects the delta between the start of the simulation and the time when the data was retrieved.

We capture two types of data as time series:
\begin{inparaenum}[(i)]
    \item sensor-based data, and
    \item trajectory data.
\end{inparaenum}
For the first type, we collect each sensor in Table~\ref{tab:overview-of-sensors} its state at every timestamp.
Various sensors provide different data types, such as binary and continuous.
For instance, on one hand, \texttt{hasABS}, \texttt{parking}, or \texttt{signal\_*} are binary and show if certain actions or features were active at the specific timestamp.
On the other hand, \texttt{rpm}, \texttt{wheelspeed}, and \texttt{steering} are examples of data fields on the continuous range.
The second type of data refers to the monitored trajectory of the SDC.
We capture the current position of the SDC at each timestamp.
The trajectories in the dataset are a time series of logged x,y, and z coordinates on the Cartesian coordinate system of the simulator.
In conjunction with the first type of data, the whole state of the SDC is preserved for analysis purposes.
Such analysis can be made for various areas to enhance SDC software research as discussed in Section~\ref{sec:implications}.

\section{Usage \& Implications}\label{sec:implications}
The availability of sensor and trajectory data enables the evaluation of testing methodologies without running expensive simulations.
Furthermore, it lowers the barrier for researchers to conduct research since no complex infrastructure setup is required for simulating SDCs.
In the following, we present various research directions for which our dataset can contribute.

\subsection{Driver AI development}
The field of SDCs is rapidly advancing, but the availability of AI models specifically designed for SDCs remains limited. Our goal is to contribute to the progress of open-source driving AI development. Regrettably, numerous researchers and practitioners face challenges due to the scarcity of computing resources required for generating training data for these models.
By providing access to our comprehensive dataset, we aim to empower the research community to overcome these barriers and facilitate the creation of prototypes for driving AIs. We believe that this collaborative effort will accelerate advancements in self-driving technology and foster innovation within the field.

\subsection{Regression testing in simulation}
Various state-of-the-art regression testing techniques for simulation-based tests use different features, such as \textit{road features}~\cite{DBLP:journals/ese/BirchlerKBGP23,DBLP:journals/tosem/BirchlerKDPP23}.
Using these features enables a cost-effective testing of SDCs.
However, using our dataset, we can identify and/or develop new features for new regression testing approaches.

Additionally, empirical evaluation of regression testing techniques for simulation-based tests is costly since tests must be executed to obtain the test outcome.
The dataset mitigates the issue of running simulations for this kind of evaluation purposes of regression testing techniques.

\subsection{CAN bus protocol support}
Most modern vehicles have different components that communicate with each other over a common shared bus system called \textit{CAN bus}.
Using the dataset consisting of sensor data, testing relevant CAN devices for the lane-keeping system is feasible with realistic CAN messages as test inputs.
Recent studies~\cite{DBLP:conf/kbse/BirchlerRKGLHKP23,DBLP:journals/ese/BirchlerKBGP23} already confirmed that the same type of data to develop testing approaches for CAN devices based on sensor data retrieved from simulation environments.

\subsection{Flakiness in simulation}
We encourage researchers to investigate with our dataset the flakiness of simulation-based tests.
In other domains, such as for unmanned aerial vehicles (UAV), we observe different behaviors between the real world and the simulated test cases and within simulations~\cite{DBLP:conf/icst/KhatiriPT23}.

These non-deterministic, i.e., flaky, behavior of simulators brings up new challenges, especially in the context of \textit{DevOps} pipelines.
Having flaky tests in a continuous integration and delivery (CI/CD) pipeline reduces the time to deployment since these tests lead to build failures.
The presence of flaky tests leads to failing tests that eventually fail the build process.
Furthermore, flaky tests do not necessarily reveal bugs in the system since only the test code might cause the flaky behavior.
In summary, flaky tests are a challenging problem for simulation-based tests as they are for traditional software systems.
With the dataset, we aim to provide initial execution data for comparison with the replicated dataset of other researchers.

\section{Remarks}
The aim of the dataset is to provide a variety of different sensor and trajectory data from 32,580 SDC simulations for research purposes.
Researchers can profit from the dataset because they do not necessarily need to execute expensive simulation-based tests to obtain the required data.

We used the \beamng simulator since several well-known SDC simulators are outdated and not maintained anymore, which is crucial for developing new SDC technologies.
\beamng is currently actively maintained by \beamnggmbh and is also widely used in academia.

For future datasets, we suggest increasing the amount of data since training data for several AI techniques is crucial.
Furthermore, we encourage researchers to replicate the dataset for analyses on research topics mentioned in Section~\ref{sec:implications}.

\section{Data availability}\label{sec:data-availability}
All the data and code to set up the \mongodb and compute the statistics is made available on \textsc{Zenodo}~\cite{sensodat-zenodo}.

\balance
\bibliographystyle{ACM-Reference-Format}
\bibliography{main}

\end{document}